\documentclass[twocolumn,amsmath,amssymb,10pt,superscriptaddress,a4paper,fleqn]{revtex4-1}
\usepackage[sort&compress]{natbib}
\usepackage[english]{babel}
\usepackage{lineno}
\usepackage{upgreek}
\usepackage{booktabs}
\usepackage{graphicx}
\usepackage{fancyhdr}
\usepackage[breaklinks=true,colorlinks=true]{hyperref} 

\newcommand{\class}[1]{\texttt{#1}}

\begin{document}
\title{Designing an upgrade of the Medley setup for light-ion production and fission cross-section measurements.}
\author{K.~Jansson}
\affiliation{Department of Physics and Astronomy, Division of Applied Nuclear Physics, Uppsala University, Sweden}
\author{C.~Gustavsson}
\affiliation{Department of Physics and Astronomy, Division of Applied Nuclear Physics, Uppsala University, Sweden}
\author{A.~Al-Adili}
\affiliation{Department of Physics and Astronomy, Division of Applied Nuclear Physics, Uppsala University, Sweden}
\author{A.~Hjalmarsson}
\affiliation{Department of Physics and Astronomy, Division of Applied Nuclear Physics, Uppsala University, Sweden}
\author{E.~Andersson-Sund\'en}
\affiliation{Department of Physics and Astronomy, Division of Applied Nuclear Physics, Uppsala University, Sweden}
\author{A.\,V.~Prokofiev}
\affiliation{Department of Physics and Astronomy, Division of Applied Nuclear Physics, Uppsala University, Sweden}
\affiliation{The Svedberg Laboratory, Uppsala University, Sweden}
\author{D.~Tarr\'io}
\affiliation{Department of Physics and Astronomy, Division of Applied Nuclear Physics, Uppsala University, Sweden}
\author{S.~Pomp}
\affiliation{Department of Physics and Astronomy, Division of Applied Nuclear Physics, Uppsala University, Sweden}

\begin{abstract}
Measurements of neutron-induced fission cross sections and light-ion production are planned in the energy range 1-40\,MeV at the upcoming Neutrons For Science (NFS) facility. In order to prepare our detector setup for the neutron beam with continuous energy spectrum, a simulation software was written using the Geant4 toolkit for both measurement situations. The neutron energy range around 20\,MeV is troublesome when it comes to the cross sections used by Geant4 since data-driven cross sections are only available below 20\,MeV but not above, where they are based on semi-empirical models. Several customisations were made to the standard classes in Geant4 in order to produce consistent results over the whole simulated energy range.

Expected uncertainties are reported for both types of measurements. The simulations have shown that a simultaneous precision measurement of the three standard cross sections H(n,n), $^{235}$U(n,f) and $^{238}$U(n,f) relative to each other is feasible using a triple layered target. As high resolution timing detectors for fission fragments we plan to use Parallel Plate Avalanche Counters (PPACs). The simulation results have put some restrictions on the design of these detectors as well as on the target design. This study suggests a fissile target no thicker than 2\,$\upmu$m (1.7\,mg/cm$^2$) and a PPAC foil thickness preferably less than 1\,$\upmu$m. We also comment on the usability of Geant4 for simulation studies of neutron reactions in this energy range.

\end{abstract}

\maketitle

\section{Introduction}
The experimental setup, called Medley \cite{Medley10y}, is planned to be used to measure neutron-induced fission cross sections and light-ion production in two different experimental campaigns at the Neutrons For Science (NFS) facility \cite{NFS2014}. Although the experiments share parts of their setups and share some experimental conditions, they also differ in some important aspects and are therefore to be regarded as two separate experiments. This paper is a continuation and refinement of the work presented in Ref.~\cite{ND13}.

Medley has previously been used with quasi-mono-energetic neutron (QMN) beams at the The Svedberg Laboratory (TSL) facility \cite{TSL} measuring light-ion production on, e.g. C, O, Si, Fe, Pb at 96\,MeV \cite{Medley96,Medley10y} and C, O, Si, Fe, Pb, Bi at 175\,MeV \cite{Medley175,Medley10y}. In the energy-region above 14\,MeV to a few tens of MeVs little data exists. Our presently intended energy range is 1-40\,MeV and thus our aim is to supplement the studies that have been performed previously, e.g. Ref.~\cite{slypen12C,benckO,benckSi,slypenFe}, with new data. The cross sections we will provide are double-differential, with respect to energy and emission angle of the respective secondary particle.

Measurements of light-ion production are important for both setting constraints on nuclear reaction models, and several applications. Neutron-induced light-ion production on silicon is important for studies of single-event effects in electronics \cite{JESD89A}, whereas the cross sections for nuclei like carbon and oxygen have applications in fast neutron therapy \cite{ntherapy}. Iron makes an interesting case for reactor applications since it is present in many construction materials. Both proton and $\alpha$ production are of concern since it can cause swelling and embrittlement of heavily irradiated materials, e.g. in fission and fusion reactors \cite{embrittlement}. For neutron dosimetry, light-ion production cross sections for several nuclei are of interest \cite{dosimetry}. Applications at these energies include spallation sources, Accelerator Driven Systems (ADS), as well as crew dosimetry for aviation \cite{aviation} and spaceflight \cite{spacecraft}.

In the second experiment, we plan to measure quasi-absolute fission cross sections, i.e. to measure relative to the elastic neutron scattering on hydrogen, which is considered to be the most accurately known standard cross section \cite{Carlson}. The cross sections of neutron-induced fission of the two most common isotopes of uranium, $^{238}$U and $^{235}$U, are fairly well known and used as standard cross sections up to neutron energies of 200\,MeV, but since they are standard cross sections they need to be known with as low uncertainties as possible. Despite past measurements of $^{238}$U(n,f) \cite{lisowski,shcherbakov,nolte1,nolte2,audouin,paradela} there are still some unresolved discrepancies above 20\,MeV \cite{Carlson}. Our aim is to determine the fission cross sections of $^{235}$U and $^{238}$U with an accuracy better than 2\%, which has not been achieved by any of the previous measurements in the energy range 20-40\,MeV. Measuring all three standard cross sections at the same time with low uncertainties will improve the situation for neutron energies above 20\,MeV.

Since we will measure angular distributions, we can also provide information on the angular anisotropy of the Fission Fragment Angular Distribution (FFAD) and its energy dependence. Especially as new multi-chance fission channels open up in the compound nucleus, the anisotropy provides information regarding the quantum numbers of the new available transition states.

The NFS facility at GANIL (Grand Acc\'el\'erateur National d'Ions Lourds), currently under construction, will provide a suitable beam for these experiments. During the construction we are upgrading the Medley setup with new detectors, of high timing resolution, for detection of fission fragments (FFs). At NFS the beam energy will be in the range 1-40\,MeV with the possibility of having a continuous neutron spectrum from the $^9$Be(d,n) reaction as well as a QMN spectrum from the $^7$Li(p,n) reaction \cite{NFS2014}. The continuous-spectrum option offers much higher flux and allows us to measure at all energies at the same time. However, with a drawback that the neutron energy must be determined using time-of-flight (ToF).

The simulations presented here aim to answer questions about how to optimise the setup and to determine achievable resolutions at NFS. The simulations are also intended for guiding the construction of new detectors and targets needed for the fission experiment. In addition the simulations produce pseudo data that can be used to develop routines for the analysis of raw data to be obtained in the experiment.

In this paper we start by describing the scope and physical setup of the two different experiments (Sect.~\ref{sec:scope}-\ref{sec:physsetup}) followed by how the setups were implemented in the simulations (Sect.~\ref{sec:simsetup}). Some non-standard solutions have been employed in the simulations which are described in Sect.~\ref{sec:fxs},~\ref{sec:exs}~and~\ref{sec:bias}. The last part of the methodology section, Sect.~\ref{sec:tof}, covers how the ToF of light ions and FFs will be measured and how they are used to deduce the neutron ToF.

\begin{figure}
\centering
\includegraphics[width=0.9\columnwidth]{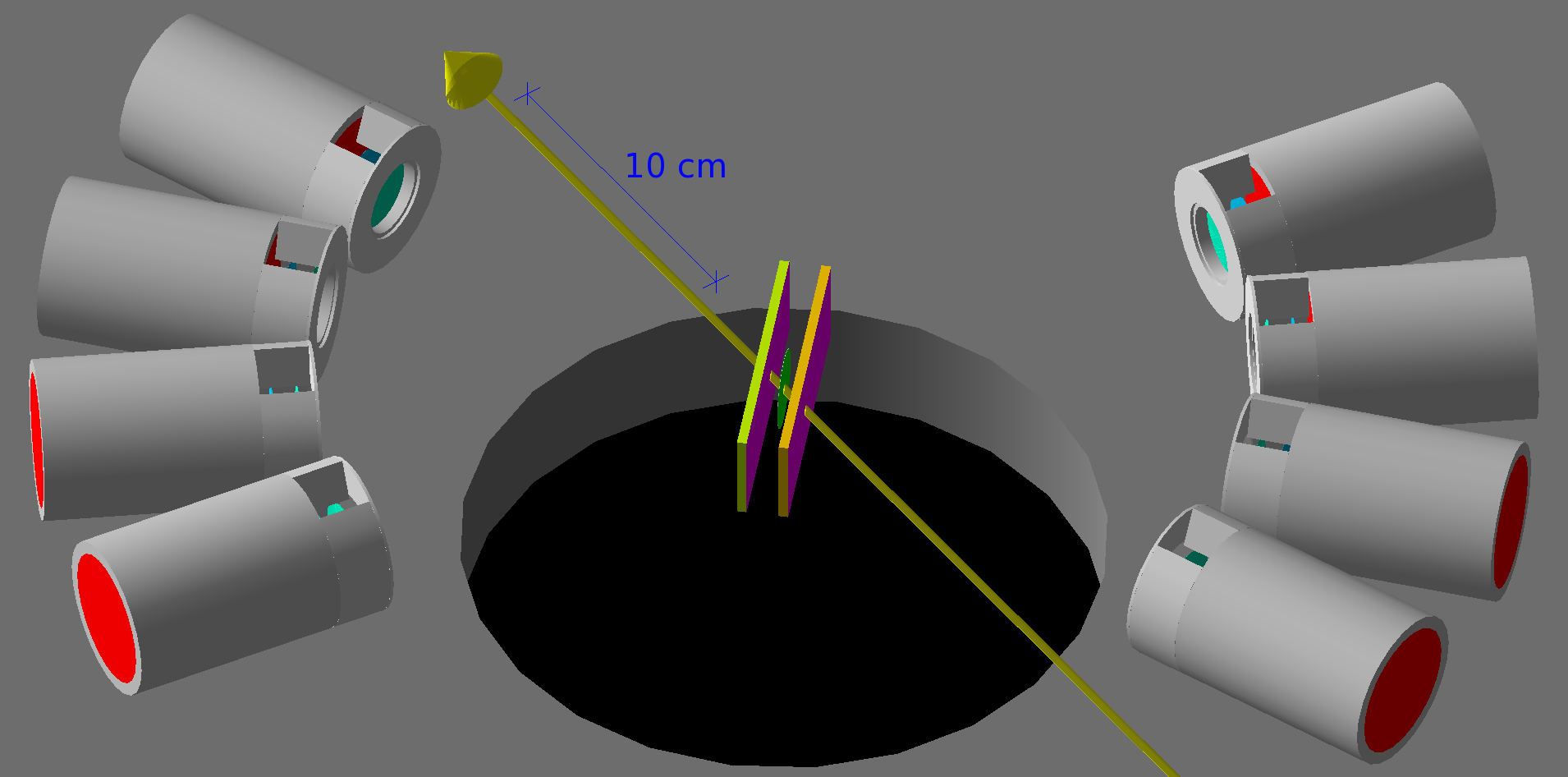}
\caption{\label{fig:simsetup}Rendered image of the simulation geometry, zoomed in on the target and detectors. The sides of the rectangular blocks facing the detectors near the centre are the Mylar foils of the PPACs. The target is centred in between the PPACs. The telescopes have two silicon detectors in the front and a CsI detector in the back. The neutrons enter from the bottom right in the $z$-direction illustrated by the arrow.}
\end{figure}

In Sect.~\ref{sec:results} we report, based on the simulation results, on the expected uncertainties and limitations of the setup. Lastly we provide a summary and outlook in Sect.~\ref{sec:conclusion}.

\section{Methodology}
The experimental setups for the two proposed experiments have been modelled and simulated with a program written using the Geant4 \cite{Geant4} toolkit. The analysis was performed using the ROOT framework \cite{Root}.

\subsection{\label{sec:scope}Scope and goals of the simulations}
The aim of the simulations, presented in this paper, is to provide information about possible setup-specific effects, e.g. due to hydrogen content in the detector material, as well as on sources of uncertainty in different scenarios. We are still developing parts of the setup and these simulations set restrictions on their design. In addition, the simulations will provide the basis for corrections needed in the data analysis.

In most aspects only relative, not absolute, simulation results are relevant for this work. Relative numbers (rather than absolute ones) will tell us where our largest setup-specific uncertainties are. For example: the anisotropy was removed from the fission model because any deviations from isotropic distributions seen in the simulations are then signs of setup-dependent effects. Similarly, exact cross section values do not greatly matter but their ratios to other cross sections do. The shape of a cross section matters since any unexpected structure appearing in the simulations must be due to the setup. However, in order for the pseudo-data to make sense, the results still need to be realistic in terms of geometry and cross sections.

In order to cover as many background effects as possible, the simulations were designed to start with the incoming neutron originating from the neutron production target. The simulations could have been simplified by starting in the target with, e.g. FFs emitted back-to-back, but details regarding, e.g. neutron-induced background in the target area, would then be lost. The background due to beam particles interacting with parts of our setup is thus simulated, but the ambient neutron background in the ToF hall is not. However, the collimator, shielding and the beam dump have been carefully designed to keep the neutron background in the time-of-flight hall as low as possible, it is several orders of magnitude lower than close to the beam line \cite{NFS2014}. A nonuniformity of the fissionable target of a few percent can be expected \cite{paradela}, but since the NFS beam profile is expected to be flat within a diameter of 3\,cm, with a variance of less than 2\% the accompanying systematic uncertainty is small since the flat part of the beam will fully cover our target.

\subsection{\label{sec:physsetup}Medley setup}
\begin{figure}
\centering
\includegraphics[width=0.5\columnwidth]{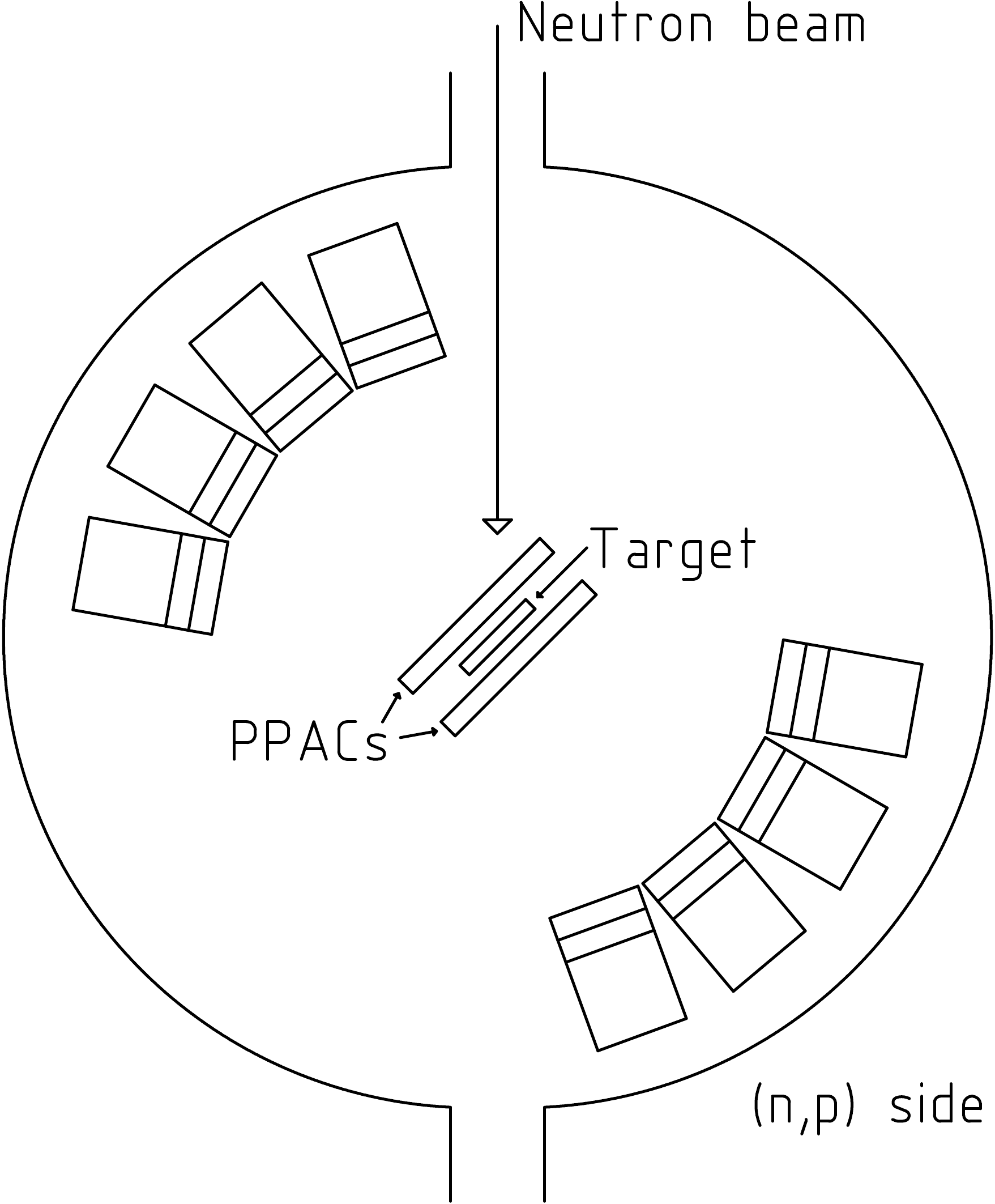}
\caption{\label{fig:medley}Schematic of the Medley setup. It consists of eight detector telescopes aligned towards the middle of the chamber where the target is put. During the fission experiment, but not when measuring light ions, two PPACs will be put close to each side of the target. The recoil protons from the elastic H(n,n) scattering can, of kinematical reasons, only be detected in the forward direction.}
\end{figure}

Medley consists of a scattering chamber about 90\,cm in diameter with a target positioned in the geometrical centre and with eight detector telescopes at adjustable distances and viewing angles (Fig.~\ref{fig:medley}). In the existing setup each telescope provides $\Delta E$-$\Delta E$-$E$ data utilising one front and one middle Si-detector as well as one CsI(Tl)-scintillator in the back. Each telescope covers about 20\,msr solid angle at 15\,cm distance from the target. One of the benefits of this setup is that the detector telescopes are attached to a rotatable plate which makes it possible for us to easily interchange the forward and backward telescopes for calibration purposes and for reducing systematic uncertainties. A more detailed description of the original setup can be found in, e.g., Ref.~\cite{Medley10y}.

The energy of an incoming neutron will be determined by using the time elapsed between the arrival of the primary beam bunch at the production target and the detection of the first reaction product. This time is a sum of the neutron ToF and the ToF of the charged secondary particle between the target and the detection point.

For the fission experiment, the Si-detectors measure the energy and arrival time of the FFs, but that alone is not enough to deduce the FF ToF due to the lack of information about the mass and the energy losses of the detected fragment. In principle we can determine the FF mass by using temporal information coming from an additional detector, e.g., the Parallel Plate Avalanche Counter (PPAC) described below, and calculate the mass from the measured energy and velocity of the FF (an expression for the FF velocity is derived in Sect.~\ref{sec:tof}). However, the separation will be rough and whether or not this is feasible at all depends on the timing resolutions of our detectors and the FF flight path length.

\begin{figure}
\centering
\includegraphics[width=0.5\columnwidth]{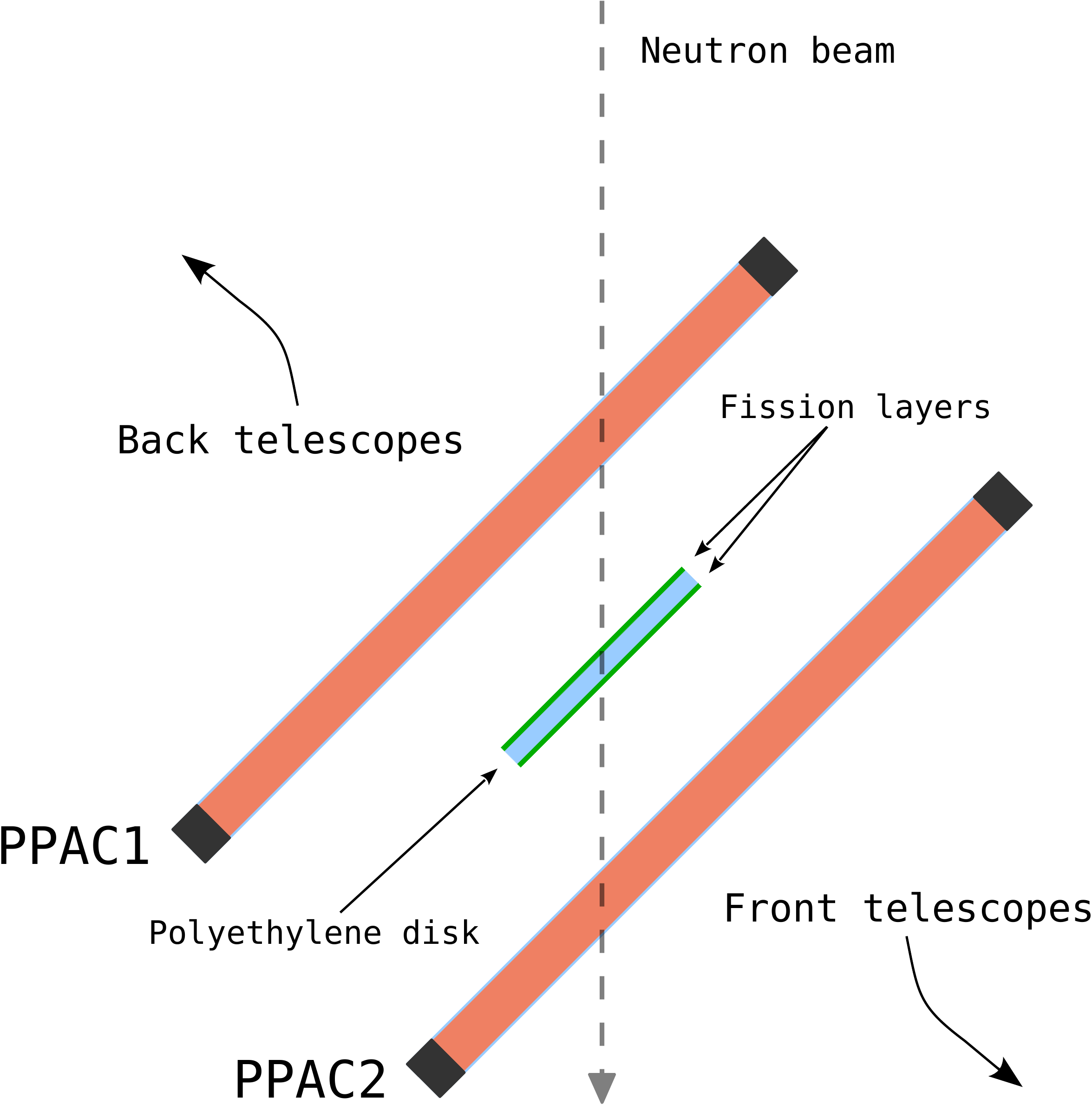}
\caption{\label{fig:target}Sketch of the area close to the target. For the fission studies we employ a layered target to measure FFs from the fissionable outer layers as well as recoil protons from the polyethylene core. The PPACs are placed close to the target in order to minimize the flightpath of the emitted FF.}
\end{figure}

Our chosen way of obtaining the ToF is to install PPACs \cite{PPACold, PPACnew, diego2014}, which have high temporal resolution, low stopping power and almost 100\% efficiency for heavy ions, close to the target (Fig.~\ref{fig:target}). The detection efficiency of these detectors improve with increasing ion mass, already oxygen ions show nearly 100\% efficiency \cite{PPACnew} and FFs are generally heavier.

The PPACs are gas-filled detectors which consist of two parallel aluminized Mylar (polyethylene terephthalate, $\left[\text C_{10}\text O_4\text H_8\right]_n$) foils. A typical gas pressure in the PPACs for this kind of experiment is a few mbar. A thin layer (much thinner than the foil thickness) of aluminium deposited on a Mylar foil allows for a high voltage to be applied over the gap between the foils. This voltage will affect the electrons that are freed upon an ionising event. As they drift towards the anode, their movement will induce a voltage which will be the signal we measure. The signal is amplified by the gas multiplication process due to the high field strength. Although the PPACs are operated in proportional mode the energy resolution is typically not better than 20\% \cite{PPACold}.

One drawback of PPACs is that the thin Mylar foils cannot handle large pressure differences on the inside of the PPAC compared to the outside. Therefore, we plan to fill the whole chamber with the same gas under the same pressure. The pressure cannot be too high since we must make sure that the FFs do not lose too much energy during their flight. Even though the Mylar foils are thin, they still introduce material containing hydrogen in the beam which will cause additional elastic scattering and thereby an increased background for the H(n,n) measurement. This background will be corrected for using the simulation results together with empty-target measurements. Note that the PPACs will not be used for the light-ion experiment in which the chamber will be evacuated.

The target for the light-ion studies will be a plain solid disk, preferably isotopically pure.

The fission studies require us to simultaneously be able to measure the H(n,n) cross section. This is done by using a layered target (Fig.~\ref{fig:target}) with fissionable material on both sides of a polyethylene disk. If the layers of both the fissionable materials and the polyethylene are made modular we can put together any mix of two outer fission layers and different thicknesses of the centre polyethylene layers (or none at all). The FF going inwards cannot escape the target but, if the fissile layer is sufficiently thin, the other FF will. This target design opens up the experimental possibility of measuring three standard cross sections at the same time if one puts $^{238}$U on one side and $^{235}$U on the other.

\subsection{\label{sec:simsetup}Simulation setup}
The simulations were performed using the Geant4 toolkit version 9.6.2. The results were recorded and analysed using classes and routines from the ROOT framework.

Whenever possible, all materials used in the simulations were NIST (National Institute of Standards and Technology) materials defined in Geant4. Custom materials were defined if a suitable NIST material was lacking.

The chamber was modelled as a stainless steel cylinder, of 90\, cm diameter, with beam pipe connectors and entrance windows of the same material. All dimensions are accurate, however, some simplifications were made. Additional entrances to the chamber for pumps etc. were neglected as well as details such as cabling and detector rails. The rotatable disc upon which the detectors are mounted was crudely approximated by yet another cylinder. A rendered image of the simulated setup can be seen in Fig.~\ref{fig:simsetup}.

The target was modelled as three layers stuck together in the middle of the chamber. For the fission study the two outer layers were made of isotopically pure $^{238}$U$_3$O$_8$ with a polyethylene layer in between. For the light-ion experiment all three layers were of the material to be studied and thus formed one uniform target.

The PPAC detectors were modelled as parallel sheets of Mylar with a thin deposit of aluminium on the side facing the other sheet. The space in between was filled with low pressure gas and so was the whole interior of the chamber. The gas used in the simulations was OctaFluoroPropane (C$_3$F$_8$) although the choice of gas to be used in the real experiment will depend on the outcome of future tests on the PPACs. The major energy losses of the FFs occur in the target and in the Mylar foils, but some losses also occur in the gas filling the chamber.

Note that in the light-ion studies the PPACs were not part of the simulation and the chamber was filled with air of low pressure (0.03\,mbar).

The detector housing was modelled to some detail but also here were details like connections and the mounting supports of the telescopes neglected. The housing material was also stainless steel whereas the silicon detectors were modelled as pure silicon with thin dead layers (80 and 225\,nm  for the front and second silicon detector respectively) consisting of SiO$_2$ \cite{Ortec}. The CsI crystals were modelled as cylinders with their backside open to the rest of the chamber. The light sensitive diodes that are attached to the backsides of the CsI crystals in the real setup were neglected in the simulation.

The simulated beam particles were created at a distance of 5\,m from the target. This distance corresponds to the expected distance between our setup and the neutron production target at the experimental site. The beam profile was assumed to be flat and of the diameter expected at NFS: 3\,cm (larger than the diameter of the target which was 2.5\,cm). The direction of the neutrons was always in the $z$-direction (towards the target). Since the beam diameter is small, compared to the distance between neutron production and the target, any angular divergence would be negligible.

The neutron energy spectrum was assumed to be flat ranging from 0.5 to 50\,MeV. The expected spectrum shape from the neutron facility is taken care of in the analysis stage where the events from the flat spectrum were weighted correspondingly. In this way we can, at any time, and without performing new simulations, change the neutron spectrum. Since the simulated energy range goes up to 50\,MeV we can decide to use a spectrum with higher energy neutrons than expected from the NFS facility. This is especially useful in future work, when the NFS spectrum has been measured, since the measured spectrum might deviate from the one we expect at the moment.

\subsection{Physics models}
For the fission simulations the physics list used was 'QGSP\textunderscore BIC\textunderscore HP' but with some modifications described below in Sect.~\ref{sec:fxs}. The fission part of the High Precision (HP) model had the option of producing FFs turned on. For the light-ion production simulations 'QGSP\textunderscore BIC' was used but with corrected cross sections as described in Sect.~\ref{sec:exs}.

\subsubsection{\label{sec:fxs}Fission cross sections}
The source code of Geant4 includes a data driven HP model for neutron reactions in the interval 0-20\,MeV. Since it is data driven, features like the increase of the fission cross section at each new chance of fission is clearly seen in the simulation results. The simulated energy range was 1-50\,MeV and a new model needs to begin at 20\,MeV giving reasonable results, as well as a smooth transition going from one model to the other. Below we will describe how the fission process was removed from one model and reimplemented in another in order to achieve a smooth transition at 20\,MeV.

The Binary cascade \cite[p.~450]{G4Physics} is a model included in the Geant4 package. It was used to describe inelastic scattering above 20\,MeV in this work. Fission is also included in this model, alongside the inelastic evaporation channels. In order to be able to modify the fission cross section so that a smooth cross section at 20\,MeV is achieved, the fission part must be separated from the inelastic part.

After the initial cascade part of the model, the subsequent evaporation is handled by a class called \class{G4Evaporation}, where fission competes with all evaporation channels. A replacement class for \class{G4Evaporation} was written. It has an identical set of channels as the original class but the probability for the fission channel was set to zero. This will give a slight increase to the probability of all the other channels since the total cross section remains the same. For this work, these small changes should have very little impact on the final results. With the fission in this process nullified, a new fission process for neutrons above 20\,MeV could be installed in its place.

We therefore wrote a new cross-section class, implementing fission cross sections for $^{235}$U, $^{238}$U and $^{232}$Th from 20\,MeV to 60\,MeV. The underlying data was taken from the IAEA neutron standards evaluations \cite{iaea} for $^{235}$U and $^{238}$U and from ENDF-VII.1 \cite{endfbvii1} for $^{232}$Th. The fission model \class{G4ParaFissionModel} was used as the replacement model together with the new custom cross-section class.

Some minor tweaks were still needed due to the usage of a parametrisation represented by the class \class{G4FissionParameters}. For high energies, e.g. above 20\,MeV, these parameters give unreasonable high probability of symmetric fission for, e.g. $^{238}$U. So to get more realistic mass distributions (see, e.g. Ref.~\cite{massdist}) the parameters corresponding to energies above 10\,MeV were fixed to the values corresponding to 10\,MeV. The resulting constant, but somewhat realistic, mass distribution does not influence the results of the simulations.

To be consistent the HP fission model was replaced by \class{G4ParaFissionModel} also for lower energies. This was implemented since the unmodified angular distribution changed abruptly at 20MeV due to G4ParaFissionModel's isotropic distribution not matching the data driven HP model's distribution. The experiment will measure the anisotropy of the cross section. However, the isotropic distribution used in the modelling is beneficial since any deviation from it, in the simulation results, can easily be seen and must be a setup dependent artefact. For the other processes below 20MeV, the HP model was kept.

\subsubsection{\label{sec:exs}Elastic cross sections}
\begin{figure}
\centering
\includegraphics[width=0.9\columnwidth]{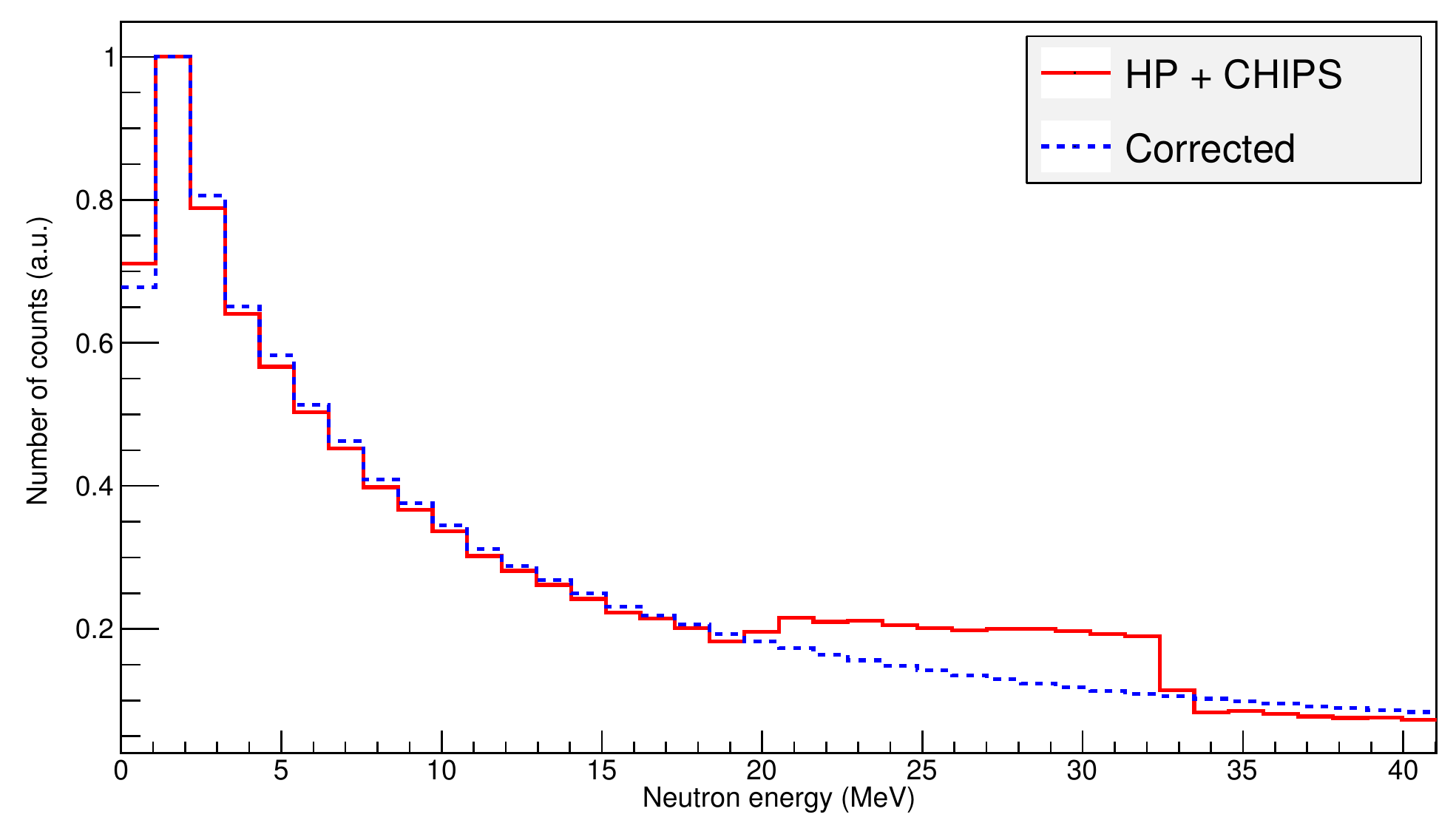}
\caption{\label{fig:elastic_xs}The histogram shows the counts of simulated recoil protons from elastic neutron scattering on hydrogen versus incident neutron energy. The results shown by the solid curve are obtained when the simulation is using cross sections provided by unmodified Geant4 classes (HP in the range 0-20\,MeV and CHIPS above 20\,MeV). After substituting the CHIPS parametrisation \cite[p.~388]{G4Physics} with additional data from evaluations \cite{jeff311} the expected smooth behaviour was restored (dashed curve).}
\end{figure}

The cross section for elastic neutron scattering on several nuclei, e.g. H, C and Si, given by the CHIPS parametrisation \cite[p.~388]{G4Physics} was found to exhibit a structure from 20 to 33\,MeV (see Fig.~\ref{fig:elastic_xs}) which deviated from the expected smooth behaviour (e.g. see evaluated data \cite{jeff311} or in the $np$ case model predictions \cite{elastic,elasticwww}). This was taken care of by a new class implemented in the same manner as for the extended fission cross sections. For most elements we kept their default cross sections, whereas carbon, hydrogen and oxygen, the main important elements in these simulations, got cross section from evaluated data \cite{jeff311} in the range 20-60\,MeV. Other elements did not occur in large enough quantities in any of the critical regions to affect the results; consequently their cross sections were not changed.

\subsection{\label{sec:bias}Biasing of neutron-induced events}
A custom made code that scaled up certain neutron cross sections was implemented, in order to save simulation time. The implementation is similar to the one described by \citeauthor{mendenhall} \cite{mendenhall}. Similar functionality has now been included in the latest Geant4 versions.

\subsection{\label{sec:tof}Time of flight}
The previous measurements using Medley have only employed QMN beams and therefore precision time-of-flight techniques were not necessary. In the case of measuring elastic scattering on hydrogen, the angle and the energy of the recoil proton were enough to identify the relevant event. Now we prepare to use a neutron beam of continuous energy spectrum and must employ ToF with high timing resolution to achieve reasonable resolution of the incoming neutron energy on an event-by-event basis. For the detection of FFs, both PPACs and silicon detectors can provide information to deduce the neutron ToF. When it comes to recoil protons or other light ions, we must rely on the silicon detectors only.

Even though the PPACs are located at a small distance, $d$, from the target, in the order of a few mm, a correction must be made in order to get the right neutron ToF when detecting FFs. Below, we will derive an expression for this correction using both the time and the energy signal from the first silicon detector in the appropriate telescope. It turns out we also need values of the average energy losses, $\Delta E_x$, in the PPAC foils and the gas volume. Since these energy losses are averages with respect to both mass and energy they can be determined offline using a $^{252}$Cf-source.

\begin{figure}
\centering
\includegraphics[width=0.9\columnwidth]{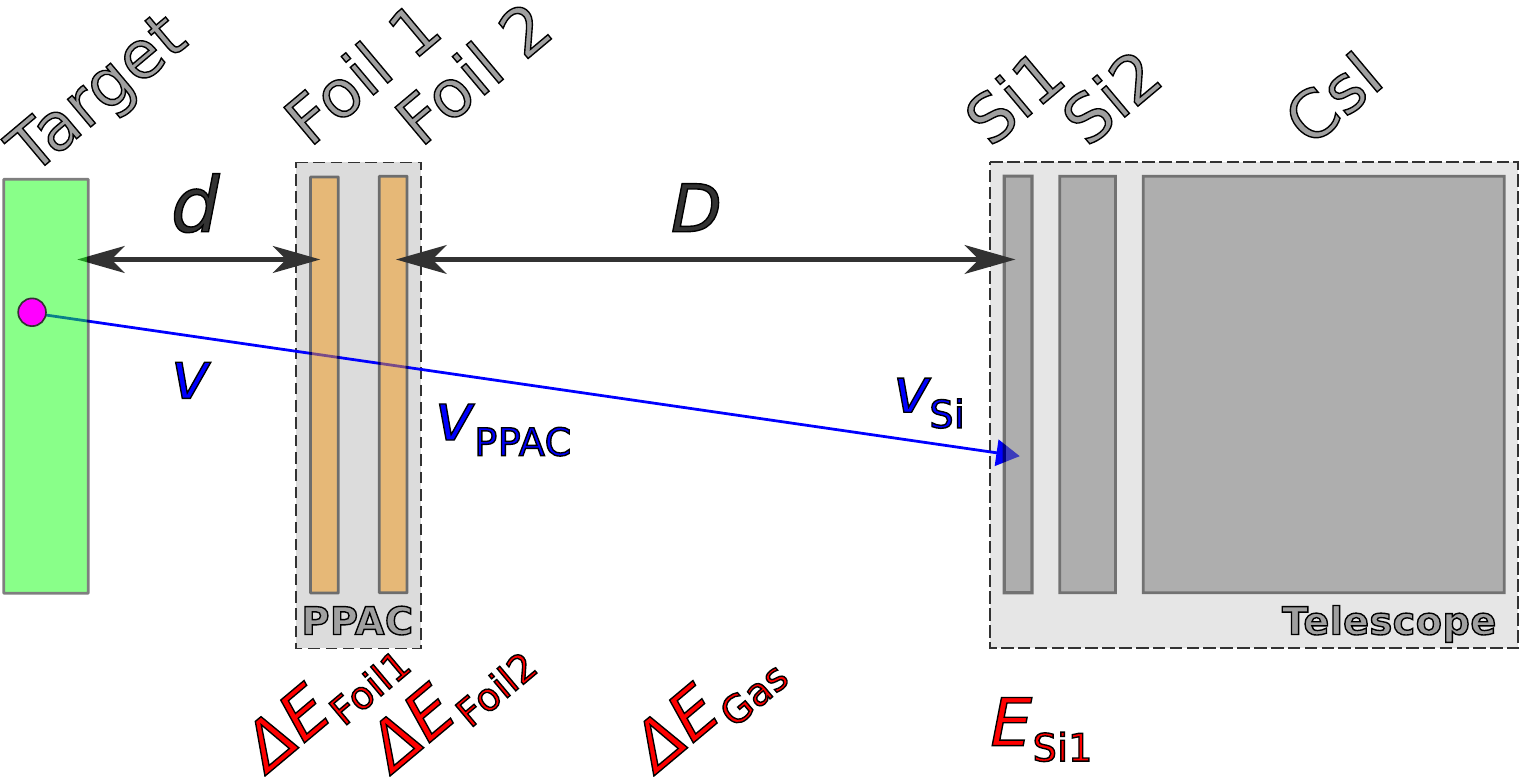}
\caption{\label{fig:tofcorr}Outline of the flight path a FF takes, travelling from the target, through one of the PPACs, into a detector telescope. The distances, velocities and the mean energy losses, used to calculate the ToF correction for the FF, are indicated in the figure.}
\end{figure}

We will find an approximate expression for the speed $v$ of the FF between the target and the PPAC. The distance, and therefore also the energy loss, between the target and the PPAC is small enough that $v$ can be regarded as constant. If $v$ can be found, the neutron ToF is given by

\begin{align}
\label{eq:v0}t_\text{ToF} = \Delta t - \frac{d}{v},
\end{align}
where $\Delta t$ is the time interval between the arrival of the primary beam bunch to the neutron production target and the passage of the FF through the PPAC. The notation is illustrated in Fig.~\ref{fig:tofcorr}.

The energy loss in the PPAC windows as well as in the gas between the PPAC and the telescope must be taken into account in order to obtain a good estimate of $v$. Since we know neither the mass nor the charge of the FF, the energy losses are estimated by a mean energy loss. The comparatively small energy losses in the gas volumes between the target and the PPAC as well as within the PPAC itself are neglected.

We start out by calculating the mean velocity of the FF between the PPAC and the telescope,

\begin{align}
\label{eq:v1} \bar v' = \frac{D}{\Delta t'} = \frac{ v_\text{PPAC} + v_\text{Si}}{2},
\end{align}
where $D$ denotes the distance between the PPAC and the front detector of the telescope, $\Delta t'$ represents the time interval beginning when the FF traverses the PPAC and ending when it hits the front telescope detector. The velocities $v_\text{PPAC}$ and $v_\text{Si}$ represent the velocity of the FF leaving the PPAC and entering the silicon detector, respectively.

A typical speed of a FF, while traversing the chamber, is $\sim$2\% of the speed of light. This is sufficiently slow so that we can use classical kinematics in order to relate fragments velocity to energy. Therefore,
\begin{align}
\label{eq:v2}\frac{E_\text{PPAC}}{E_\text{Si1}} = \frac{E_\text{Si} + \Delta E_\text{gas}}{E_\text{Si1}} = \left(\frac{v_\text{PPAC}}{v_\text{Si}}\right)^2,
\end{align}

where the energy loss of the FF in the object $x$ is denoted by $\Delta E_x$. Combining Eqs.~\ref{eq:v1}-\ref{eq:v2} yields
\begin{align}
\label{eq:v3}v_\text{Si} = \frac{2\bar v'}{1+\sqrt{\frac{E_\text{Si1} + \Delta E_\text{gas}}{E_\text{Si1}}}}.
\end{align}

Considering the fragment energy before the fragment reaches the PPAC we get, similarly to Eq.~\ref{eq:v2}
\begin{align}
\label{eq:v4}\frac{E_\text{Si1} + \Delta E_\text{gas} + \Delta E_\text{Foil\,1} + \Delta E_\text{Foil\,2}}{E_\text{Si1}} = \left(\frac{v}{v_\text{Si}}\right)^2.
\end{align}

Eq.~\ref{eq:v4} allows us to calculate $v$ with the help of Eqs.~\ref{eq:v1} and\,\ref{eq:v3}:
\begin{align}
\label{eq:v5}v = v_\text{Si} \sqrt{\frac{E_\text{Si1} + \Delta E_\text{gas} + \Delta E_\text{Foil\,1} + \Delta E_\text{Foil\,2}}{E_\text{Si1}}}.
\end{align}

Although one might think that it is crude to use only average values in the corrections outlined in Eqs.~\ref{eq:v0}-\ref{eq:v5}, this method has worked well when it was applied to simulated pseudo data as will be demonstrated in Sect.~\ref{sec:res}.

The ToF for recoil proton or light-ion production events are more straightforward. Since we determine the ion species and energy through the $\Delta E$-$E$ technique we can calculate the ToF, provided that the time resolution in the detectors are good enough. The time resolutions needed is discussed in Sect.~\ref{sec:res}.

\section{\label{sec:results}Results}
\subsection{\label{sec:res}Time resolution and neutron energy uncertainties}
\begin{figure}
\centering
\includegraphics[width=\columnwidth]{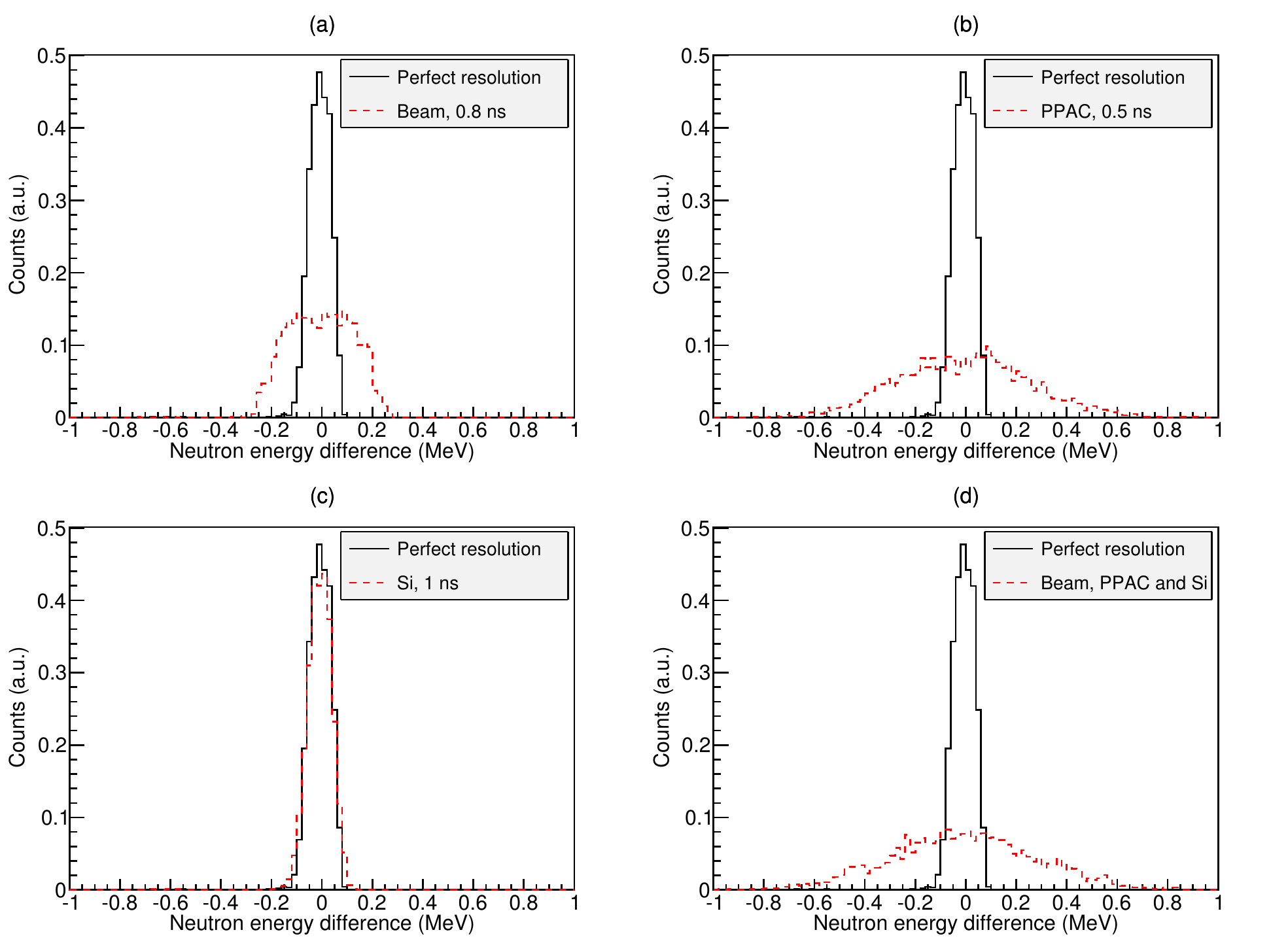}
\caption{\label{fig:tof_energy}Distributions of the difference between the simulated (true) neutron energy and the neutron energy calculated from the simulated ToF. The solid line histograms show the distributions with perfect detector resolution and no temporal structure in the primary beam. Note that even with perfect detector resolution we still have effects due to different FF masses, variations in the paths taken by the FFs as well as the angular acceptance of the silicon detectors. A few different scenarios are shown. In (a) one sees the effect of adding temporal structure (FWHM 0.8\,ns) to the primary beam. In (b) only the PPAC resolution ($\sigma_\text{PPAC}=0.5$\,ns) is taken into account and similarly in (c) only the effect of the silicon detector's resolution ($\sigma_\text{Si}=1$\,ns) is present. Finally, in (d) all the above mentioned effects are taken into account. The histograms are based on $\sim$2\,600 events, all with neutron energies close to 20\,MeV ($\pm 1$\,MeV).}
\end{figure}

We have investigated the ToF distribution of FF events, taking into account the temporal resolutions of the detectors as well as the duration of the primary beam bunch. For each event we compared the true incident neutron energy with the one deduced from the ToF as outlined in Sect.~\ref{sec:tof}. The time spread coming from the duration of the beam bunch was based on the expected properties of the NFS facility \cite{NFSTech} where a FWHM of about 0.8\,ns is expected for 40-MeV deuteron bunches (using a buncher). In Fig.~\ref{fig:tof_energy} one sees the difference between the deduced and the true value for a few different scenarios at a neutron energy of $\sim$20\,MeV (within $\pm 1$\,MeV). Some intrinsic limits on the achievable resolution, due to e.g. the different speeds and paths of the FFs, are unavoidable. Without any added resolution effects the uncertainty (RMS) is 0.05\,MeV. If the effects of the primary beam's time structure and the detector resolutions are added, the uncertainty in the neutron energy determination increases.

The rise time of a preamplified silicon detector signal is typically less that 10\,ns \cite[p. 395]{knoll}, so by digitizing and interpolating the signal, or using equivalent analogue equipment, it is reasonable to assume that we can reach at least a 1\,ns resolution. The time resolution of PPACs can be lower than 200\,ps even for light particles \cite{breskin}. We have based our investigation on the assumed resolutions of 0.5\,ns and 1\,ns for the PPAC and silicon detector respectively, and compared the different contributions to the overall neutron energy resolution. The RMS values of the different distributions are tabulated in Table~\ref{tab:tof_energy} but also depicted in Fig.~\ref{fig:tof_energy}. No dependence on the gas pressure used in the chamber has been detected, once again indicating that the correction previously outlined in Eqs.~\ref{eq:v0}-\ref{eq:v5} works as intended.

\begin{table}
\centering
\caption{\label{tab:tof_energy}The RMS. values of the distribution found in Fig.~\ref{fig:tof_energy} where different resolution effects are taken into account. The beam resolution effect is due to the temporal structure of the primary beam (FWHM 0.8\,ns). The detector resolutions of the PPAC and Si detectors were taken as $\sigma_\text{PPAC}=0.5$\,ns and $\sigma_\text{Si}=1$\,ns respectively.}
\begin{tabular}{lc}
\toprule
Included resolution sources & RMS (MeV) \\
\midrule
None & 0.05 \\
Beam & 0.13 \\
PPAC & 0.26 \\
Si   & 0.05 \\
Beam, PPAC, Si & 0.29 \\
\bottomrule
\end{tabular}
\end{table}

The relative neutron energy uncertainty of the FF events is dominated by the PPAC time resolution with significant contributions at higher energies from the primary beam bunch duration. If a PPAC resolution of 0.1\,ns could be reached, the major uncertainty regarding the neutron energy would be due to the beam. With a time resolution of 1\,ns, the silicon detectors would not contribute significantly to the neutron energy uncertainty. These conclusions remain valid even at quite large variations (up to 10\%) of the mean energy losses used in the correction outlined in Eqs.~\ref{eq:v0}-\ref{eq:v5}.

\begin{figure}
\centering
\includegraphics[width=\columnwidth]{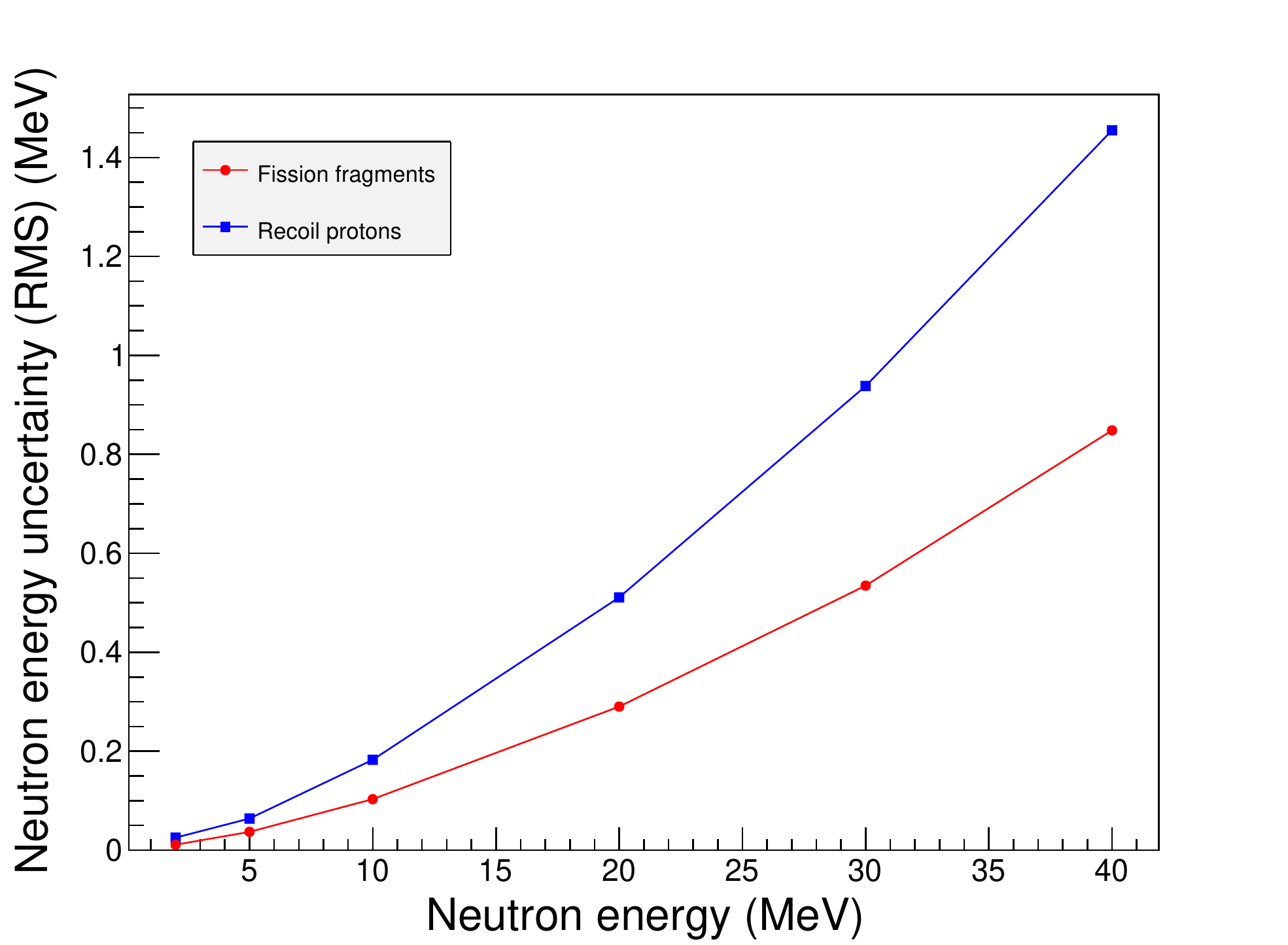}
\caption{\label{fig:neres}The plot shows calculations of the neutron energy resolution as a function of neutron energy. For the detectors we assumed a resolution in the PPACs and front silicon detector of 0.5\,ns and 1\,ns respectively. The resolution is different for detection of FFs and recoil protons since the detection techniques and the respective ToF analyses are different. A FWHM of 0.8\,ns has been assumed for the temporal structure of the primary beam. The lines are mere guides for the eye.}
\end{figure}

In Fig.~\ref{fig:neres}, we show the simulation-based uncertainty of the deduced neutron energy versus the true neutron energy, for FF and recoil-proton events. Over the whole energy range we expect a resolution not worse than 1.8\% for FF events. The worse time resolution of the silicon detector compared to the PPAC emerges in the recoil proton case as a higher uncertainty of the neutron energy. It is still no more than 3\% over the full range.

The neutron energy uncertainty, compared to the overall uncertainty of the cross section, depends on the shape of the cross-section curve at a given neutron energy. The critical energies are those where the cross section changes rapidly with energy within the range of an energy bin. The H(n,n) cross section is slowly varying with energy over most of the studied energy range, which results in insignificant errors due to the limited neutron energy resolution. Also the fission cross section is quite flat except at energies where new chances of fission open up \cite{multichance}. Therefore, the neutron energy uncertainty is not expected to give a major contribution to the total uncertainty.

Since the energy of the recoil protons is proportional to $\cos \theta$, where $\theta$ is the polar angle with respect to the beam direction, the acceptance angle of the telescopes also affect the uncertainties. The silicon detectors are 23.9\,mm in diameter and cover about 9.1$^\circ$ of the polar angle interval at a distance of 15\,cm from the target. However, due to the circular shape of the detector the aperture is better quantified by the angular RMS, $\sqrt{\langle\Delta \theta^2\rangle}\approx 2.3^\circ$, where $\Delta \theta$ is the distance from the centre of the detector. Within the angular range of one of our silicon detectors, the angular-differential H(n,n) cross section for our smallest measured laboratory angle, 20$^\circ$, does not change more than $\sim$2\% and it does so in a smooth predictable manner. Therefore, we do not expect the silicon detectors' angular acceptance to influence the uncertainty of our measurement significantly.

The light-ion production measurement will experience similar uncertainties as the measurement of recoil protons of the fission experiment. The main difference is that when a reaction product from light-ion production is detected, the specific reaction that produced it is unknown. So generally, the light-ion production lacks of the strict kinematic constraint between the particle energy and the emission angle that is found in elastic scattering. Comparing the uncertainties associated with the detection of protons with those for $\alpha$-particle detection, we see a similar timing resolution for $\alpha$-particles.

Nuclear reactions in the CsI detectors can cause the incoming light ions not to deposit their full energy in the detector. This effect has been studied before and was shown to be less than 2\% for 40\,MeV protons \cite{hayashi}. It was also found that the effect can be precisely reproduced by simulations and therefore appropriate corrections can be made.

Light ions, with energies of only a few MeV, lose a significant part of their energy in a hundreds of $\upmu$m thick target. Corrections for this energy loss can be made, e.g. with the method described in Ref.~\cite{tcorr}, with an additional contribution to the overall uncertainty.

\subsection{\label{sec:targetfoil}Target and foil thicknesses}
For the fission experiment some limits on the target thickness must be put. Due to the thick middle layer of polyethylene, a FF emitted inwards will always be lost. The fission material deposit must be thin in order to assure that a fragment emitted outwards, towards one of the eight telescopes, will always be detected. The energy losses in the PPACs and in the gas that fills the whole chamber also need to be taken into account.

Moreover, it is not enough, only for the fragment to reach the silicon detector, it must also deposit enough energy in the detector. To distinguish FFs from $\alpha$-particles one can use the pulse height of the PPAC signal \cite{PPACold}. However, some overlap between the FF and $\alpha$-particle spectrum is possible \cite{Diegothesis}. One can also differentiate between FFs and $\alpha$-particles using the front silicon detector in the telescopes. For a 25-$\upmu$m thick detector the maximum energy deposited by an $\alpha$-particle will be about 5\,MeV, whereas higher-energy $\alpha$-particles will punch through. For a thicker detector of 50\,$\upmu$m the corresponding punch-through energy is about 8\,MeV. Therefore, a FF can be distinguished from an $\alpha$-particle if the former deposits more than 5 and 8\,MeV respectively. To fulfil this requirement the total energy loss of the fragment must be kept low enough.

\begin{figure}
\includegraphics[width=\columnwidth]{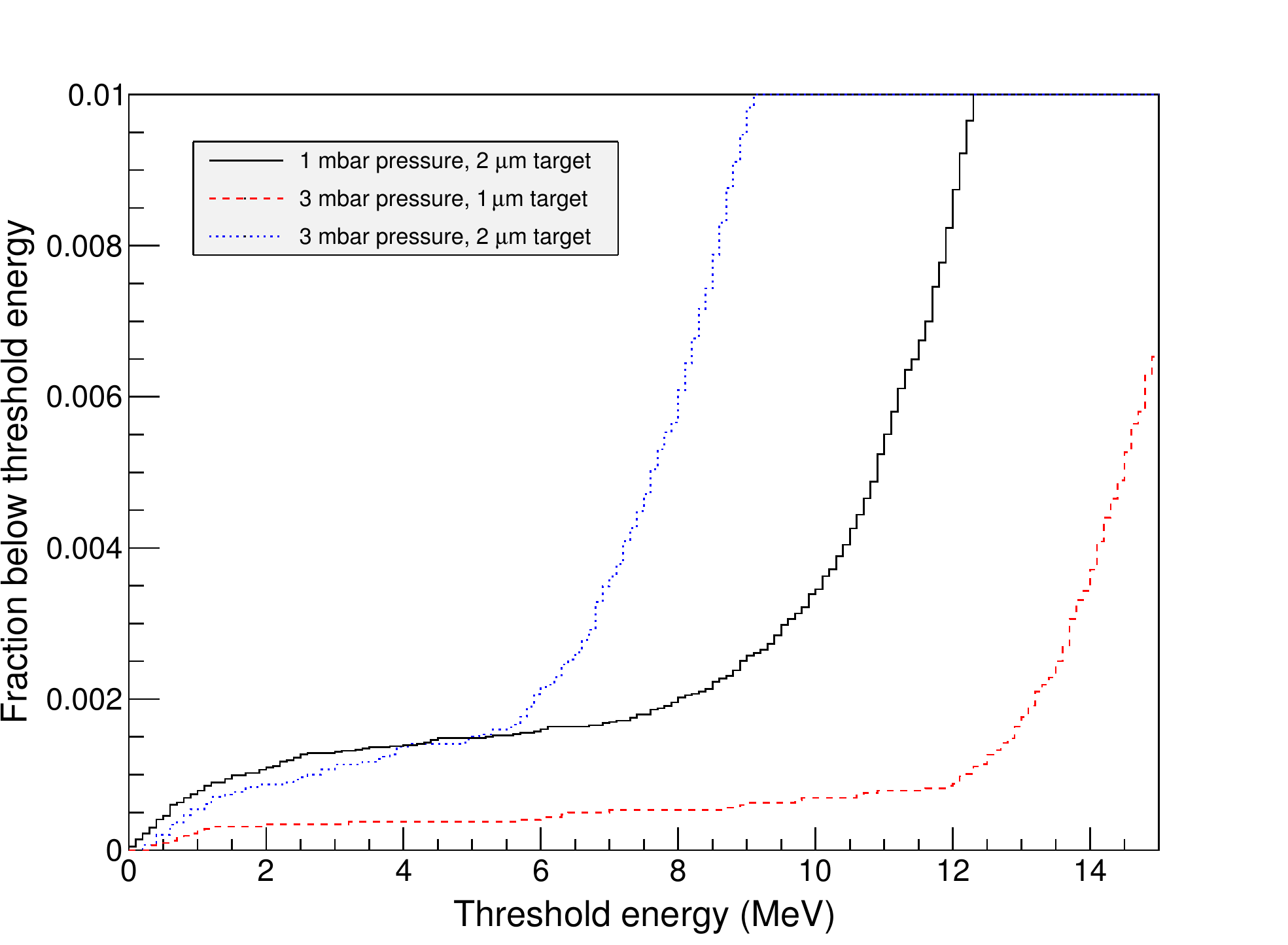}
\caption{\label{fig:ffesi}The fraction of FFs that reach the front silicon detector with energy lower than the threshold indicated on the $x$-axis. Most of the fragment's kinetic energy is lost while escaping the target and penetrating the PPAC foils as well as traversing the gas towards the detector telescopes. The PPAC foil thickness is kept at a minimum, 1\,$\upmu$m. If the gas pressure is kept at 1\,mbar and the target is 2\,$\upmu$m (1.7\,mg/cm$^2$) thick, about 99.8\% of the FFs are detected with energies higher than 8\,MeV. To detect more FFs above the $\alpha$-punch-through energy, the target thickness can be decreased to 1\,$\upmu$m (0.83\,mg/cm$^2$) allowing almost every FF to reach the silicon detector with a kinetic energy above 8\,MeV.}
\end{figure}

If the Mylar foil, that constitutes the PPAC windows, is 1\,$\upmu$m thick, the FFs lose in average about 8\, MeV per penetrated foil. But, if a fission occurs deep inside the target, a significant part of the fragment's kinetic energy will be lost before the FF even reaches the PPAC. The fraction of FFs reaching the detector telescopes with at least a certain kinetic energy is shown in Fig.~\ref{fig:ffesi}. Up to a deposited energy of about 6\,MeV the curves corresponding to a target thickness of 2\,$\upmu$m (1.7\,mg/cm$^2$) have overlapping error bars, but after that they diverge quickly.

The Mylar window foil thickness must be kept as thin as possible. Alas, we have not found a supplier of Mylar foils thinner than 1\,$\upmu$m. If a gas pressure of 3\,mbar is used together with a target thickness of 2\,$\upmu$m (1.7\,mg/cm$^2$), 99.4\% of the FFs are detected with a higher energy than 8\,MeV (the $\alpha$-punch-through energy of a 50\,$\upmu$m Si detector). With a gas pressure decreased to 1\,mbar the ratio of FFs increases to 99.8\%. If the gas pressure is to be lowered from 3\,mbar to 1\,mbar, in order to keep the same reduced field strength, one must either increase the distance between the parallel plates, with reduced timing resolution as a consequence, or decrease the applied voltage, which will decrease the multiplication factor.

When decreasing the target thickness to 1\,$\upmu$m (0.83\,mg/cm$^2$), but keeping the gas pressure at 3\,mbar, 99.9\% of the fragments deposit at least 12\,MeV in the silicon detector. Though, decreasing the target thickness comes with the cost of a decreased event rate.

Considering our goal of a total uncertainty of less than 2\% for the angular-integrated cross section, it can be tolerable that a 0.2\% of the FFs cannot be distinguished from $\alpha$-particles. Higher ratios, in the order of 1\%, cannot be accepted without a very reliable correction method. Using 25\,$\upmu$m, rather than 50\,$\upmu$m, thick silicon detectors eases the situation since the punch-through energy then becomes lower.

In addition to $\alpha$-particles some Li-particles will also be produced. The higher stopping power of Li-particles opens for the possibility to misinterpret them as FFs. In the simulations, Li-particles are more than an order of magnitude less frequent than FFs. Also, most of the Li-particles are produced with energies below the $\alpha$-punch-through threshold which makes it impossible for them to deposit more energy than that in the silicon detector. Both of these effects contribute to the fact that the risk of interpreting a Li-particle as a FF is negligible and no such events were observed in the simulations.

The current setup has 50-$\upmu$m thick front silicon detectors, but also 25-$\upmu$m thick ones are being considered, in order to lower the $\alpha$-particle threshold. A possibility which has not been investigated by simulations is to distinguish particle types by pulse shape analysis (PSA). A working PSA would relax the requirements on the thicknesses, but to be used optimally it requires digital sampling of the pulse shapes in the front silicon detector.

\subsection{PPAC hydrogen content}
During fission measurements, the PPACs will be placed in the neutron beam. The presence of hydrogen in the Mylar foils adds additional scattering centres for the H(n,n) reaction. If not corrected for, this leads to an overestimation of the elastic scattering cross section. The atomic ratio of the amount of hydrogen in the target to the amount of hydrogen in the PPACs can be estimated, if the corresponding thicknesses are known. Assuming a polyethylene thickness of 100\,$\upmu$m and a total Mylar thickness of 4\,$\upmu$m (each of the two PPACs has two windows) we arrive at a ratio of about 2\%. The exact correction will also depend on the final geometrical design of the PPACs as well as the beam diameter. Since the target might partly shadow the telescopes from some of the extra recoil protons, the correction will be based on a combination of empty-target runs and simulations.

\subsection{\label{sec:anisotropy}Fission Fragment Angular Distribution (FFAD)}
Since we only detect FFs at eight specific polar angles $\theta$, we need to estimate the total fission cross section by interpolation with subsequent angular integration. The FFAD, for each neutron energy, is transformed to the centre-of-mass system by calculating the linear momentum transfer assuming FFs with average mass and kinetic energy. The interpolation is performed by fitting the centre-of-mass FFADs with Legendre polynomials. Previous measurements \cite{diego2014} show that we need to use at least 4\textsuperscript{th}-order polynomials (only even orders contribute since the centre-of-mass FFADs are symmetric in front and backwards angles). Assuming that the data can be well represented by the polynomial we expect the uncertainty for the angular-integrated cross section calculated from the fitted Legendre polynomials to be close to the statistical uncertainty in all data points combined. By rotating the telescope setup we can cover additional angles if needed. As a by-product we can provide angular anisotropy data in addition to the cross sections.

No deviation from an isotropic FFAD could be observed in the simulation results, indicating that there are no significant setup dependent effects affecting the FFAD. In reality though, we expect to see an energy dependent angular anisotropy.

As mentioned in Sect.~\ref{sec:physsetup}, we can roughly separate the light and heavy FF masses by calculating the FF velocity using Eq.~\ref{eq:v3}. By relating the velocity to the kinetic energy measured in the silicon detector we can in principle calculate the mass. To achieve separation of the distributions of the light and the heavy FFs, based on the timing resolutions used in Sect.~\ref{sec:res}, we would need to increase the FF flight path length. Our default and closest possible distance between the telescopes and the target is 15\,cm, but it can be increased to 30\,cm at the cost of losing 75\% solid angle coverage. Distinguishing the FFs from symmetric fission from the light and heavy FFs, opens up the possibility of measuring the angular anisotropy for both the symmetric and asymmetric fission modes.

\section{\label{sec:conclusion}Discussion}
The use of the Medley setup to measure fission cross sections and light-ion production at the upcoming NFS facility has been discussed. The consequences of having a continuous neutron spectrum and the necessity of including fast PPAC detectors in the setup for the fission measurements have been investigated. To estimate uncertainties and resolutions as well as the design of the PPACs, Geant4 simulations were performed.

The neutron energy region of this work, 1-50\,MeV, can be challenging to simulate. Many reaction models rely on data or semi-empirical formulas that, in turn, are based on measured nuclear data. The energy region above 14\,MeV is often less investigated due to the scarcity of suitable neutron sources. In effect, many evaluated nuclear data libraries stop at 20\,MeV and so does the data driven neutron model (HP) in Geant4. At this energy, the simulation must make a smooth transition into a new model. For fission cross-sections, the quality of the non-data-driven models were found to be insufficient. In addition, an artefact in the standard model for the cross section of elastic neutron scattering was found. In order to resolve these problems, customised data-driven models were implemented as described in Sect.~\ref{sec:fxs} and~\ref{sec:exs}. If future Geant4 versions include either an extension of the HP neutron models to higher energies, or just a more easy-to-use interface to add one's own data for any reaction, the situation would be improved greatly for the general user.

For the detection of light ions, the limiting factor is the timing resolution of the silicon detectors, whereas it is the PPAC timing resolution for FFs. For the detectors to meet the requirements of the experiment, it was found in Sect.~\ref{sec:res} that we must have timing resolutions better than 1\,ns and 0.5\,ns for the silicon and PPAC detectors, respectively. If we succeed in achieving a PPAC timing resolution as good as 0.1\,ns, the temporal characteristics of the primary accelerated beam will become the limiting factor. We expect an uncertainty in the neutron energy, determined by the ToF techniques, to be 3-4\% for light ions and less than 2\% for FFs. The influence of the neutron energy uncertainty on the measured data will be negligible except at energies where the cross-section varies significantly within an energy bin. For the fission studies this means that an absence of sharp changes in the cross section leaves us with a corresponding uncertainty, due to the neutron energy, of far less than 1\%, which is less than the statistical uncertainty we are aiming at.

For each fission event we need to detect at least one fragment, which puts a limit on the target thickness. It was found in Sect.~\ref{sec:targetfoil} (illustrated in Fig.~\ref{fig:ffesi}) that a thickness thinner than $\sim$2$\,\upmu$m ($\sim$1.7\,mg/cm$^2$) is required for the FFs to escape a $^{238}$U$_3$O$_8$ target with enough kinetic energy given a gas pressure of 1\,mbar. A thinner target allows for a higher gas pressure. In order to clearly differentiate between $\alpha$-particles and FFs using the energy depositions in the silicon detectors we need to balance the target thickness (which determines our total count rate) against the gas pressure (which influences the performance of the PPACs). In order to keep our nominal gas pressure of 3\,mbar, we must choose a thinner target.

The expected neutron flux in Ref.~\cite{NFS2014} varies with as much as an order of magnitude depending on the neutron energy, but the average neutron flux is in the order of 2\textsc{e}6\,n/MeV/cm$^2$/s. With this we expect a fission count rate in the order of 1\,kHz per $\upmu$m target thickness. The count rate in the detector telescopes will be much lower due to the small solid-angle coverage, around 3\,Hz per $\upmu$m target thickness. Considering the high neutron flux we expect to get a statistical uncertainty below 1\% using 1-MeV bins within a few hours of beam time, so we do not expect the statistical uncertainty to be a large part of our total uncertainty. If bins as large as 1\,MeV are used, there will be regions where the cross section changes significantly within a bin, which will affect the uncertainty at these specific energies. However, this effect is largest at about 1.5\,MeV, close to the opening of the fission channel, and for higher energies the cross-section variation within a 1-MeV bin is less than 10\%. Since we have a rather small neutron energy uncertainty, the expected uncertainty due to the binning is in the order of per mille.

Aided by this work, we are going to finalise the design of the PPACs and the targets for the fission experiment. It is critical to reduce the PPAC foil thickness as much as possible since it enhances the discrimination between FFs and $\alpha$-particles. In this way, the hydrogen content, originating from non-target objects, in the beam path would also be reduced, minimising the needed corrections. Silicon detectors thinner than 50\,$\upmu$m do not seem to be necessary, if the target thickness is reduced to 1\,$\upmu$m (0.83\,mg/cm$^2$) or less. Together with better information on the neutron beam profile and the final design of the PPACs these simulations will be the basis for the correction procedure in order to compensate for the extra H(n,n) scattering in the PPACs.

The FFADs will need to be estimated in order to calculate the total fission cross section. By doing so, we will, in addition to the cross section, provide FFADs for each measured neutron energy bin. Better experimental FFAD data, to test theoretical models against, can improve the description of the fission phenomena.

We have shown that the Medley setup, after being upgraded to cope with a continuous-energy neutron beam, will be able to measure both light-ion production and fission cross-sections at intermediate energies (1-40\,MeV) at the upcoming NFS facility. For the fission studies we have investigated and discussed several sources of uncertainties: statistical, target and beam uniformity, neutron energy resolution, detection efficiency, $\alpha$-particle discrimination as well as additional H(n,n) scattering centres due to the PPACs. None of these sources have been able to pose a critical challenge to our goal of reaching a 2\% uncertainty for our fission cross-section measurements.

\section*{Acknowledgements}
This work was supported by the Swedish Research Council.

\bibliography{jansson_et_al_-_medley_upgrade}

\end{document}